\documentstyle[12pt]{article}
\newcommand{\NP}[1]{Nucl. \ Phys.}
\newcommand{\PL}[1]{Phys. \ Lett.}
\newcommand{\p}[1]{\partial}
\newcommand{\PRL}[1]{Phys.\ Rev.\ Lett. }
\newcommand{\MPL}[1] { Mod. Phys. Lett. }
\newcommand{\IJMP}[1] { Int. J. Mod. Phys. }
\begin{document}

\title{
Planckian Energy Scattering of D-branes\\
 and\\
M(atrix) Theory in Curved Space}
\author{
I.V.Volovich \\Steklov Mathematical Institute\\
Gubkin St.8, GSP-1, 117966, Moscow, Russia}
\date {$~$}
\maketitle
\begin {abstract}
We argue that black $p$-branes will occur in the collision of D$0$-branes at
Planckian energies.  This extents the Amati, Ciafaloni and Veneziano and 't
Hooft conjecture that black holes occur in the collision of two light
particles at Planckian energies. We discuss a possible scenario for such a
process by using colliding plane gravitational waves.  D-branes in the
presence of black holes are discussed. M(atrix) theory and  matrix string in
curved space are considered.  A violation of quantum coherence in M(atrix)
theory is noticed.

\end {abstract}
\section{Introduction}

It was suggested \cite{BFSS} that the large $N$ limit
of the dimensional reduction of the ten dimensional $U(N)$ supersymmetric
Yang-Mills theory to one dimension  might
be interpreted as eleven dimensional M-theory
in light cone gauge. The reduction of the supersymmetric Yang-Mills
theory to two dimensions was interpreted as matrix string theory
\cite{Motl}. This is based on the observation \cite{Wit} that the effective
low energy theory of $N$ coincident parallel Dirichlet  branes is
described by the dimensional reduction of the $U(N)$ supersummetric
Yang-Mills theory.

M(atrix) theory \cite{BFSS} is described by the following
Lagrangian
\begin{eqnarray}
\label{I1}
L=\frac{1}{2}tr[\dot{ Y}^{i}\dot{Y}^i+\frac{1}{2}[Y^i,Y^j]^2
+2\theta ^T\dot{\theta}+2\theta ^T\gamma_i[\theta ,Y^i]]
\end{eqnarray}
where $Y^i$ are Hermitian $N\times N$  matrices while $\theta$
is a 16-component fermionic spinor each component of which
is an Hermitian $N\times N$ matrix and $i,j=1,...,9$.

In \cite{DFS,DKPS} the interaction between D-branes in the non-relativistic
approximation has been considered. It was found \cite{DKPS}
that supergravity
is valid at distances greater than the string scale, while
the description in terms of gauge theory
in flat space is valid in the sub-stringy
domain.

The application of the flat background is justified
if the curvature is small. However if one has
D-branes in the presence of a black hole we have to take into account
the non-trivial background. The low
energy bulk theory of $D$-branes is the supersymmetric
Yang-Mills theory coupling with supergravity in ten dimensions.
In this note we
 discuss the interpretation of
the dimensional reduction of this theory
as M-theory (if $p=0$) or matrix string (if $p=1$) in curved background.
We will argue that
 (black) $p$-branes
will occur in the collision of D$0$-branes at Planckian energies.

\section{M(atrix) theory in curved space}

The low energy effective theory of D-branes in Minkowski spacetime
is given by the dimensional reduction of the supersymmetric
gauge theory in the ten
dimensional Minkowski
spacetime \cite{Wit}. If one has D-branes
in Minkowski spacetime but in  curved coordinates we have to
start from the supersymmetric Yang-Mills theory in curved coordinates.
Then we get  a version of the M(atrix) theory Lagrangian
(\ref{I1}) in the curved coordinates.  If one has D-branes in a curved
spacetime, for instance D-branes in the presence of black hole then it is
natural to expect that the low energy effective theory will be given by the
dimensional reduction of the supersymmetric gauge theory coupling with
supergravity in the ten dimensional curved spacetime.

Let us consider the Yang-Mills theory in the $D$-dimensional
space-time with
the metric $g_{MN}$. The action is
\begin{eqnarray}
\label{M1}
I=-\frac{1}{4}tr\int d^Dx\sqrt g F_{MN}F_{PQ}g^{MP}g^{NQ}
\end{eqnarray}
where $F_{MN}=i[D_M,D_N],~~ D_M=\nabla_M -iA_M$. Let $\gamma :X^M=
X^M(\sigma),~
\sigma=(\sigma_0,...,\sigma_p)$ be a $p+1$-dimensional submanifold
($p$-brane) and let us consider the dimensional reduction to $\gamma$.
One has $A_M=(A_{\alpha},~ Y_i), \alpha =0,...,p;~ i=p+1,...,D-1$
and $F_{MN}=(F_{\alpha\beta},F_{\alpha i},F_{ij}),~
g^{MN}=(g^{\alpha\beta},g^{\alpha i},g^{ij})$. The Lagrangian
is
\begin{eqnarray}
\label{M2}
L=-\frac{1}{2}tr[D_{\alpha}Y_iD_{\beta}Y_jg^{\alpha\beta}g^{ij}
-\frac{1}{2}[Y_i,Y_j][Y_m,Y_n]g^{im}g^{jn}+...]
\end{eqnarray}
Here $Y_i=Y_i(X^P(\sigma)),~g_{MN}=g_{MN}(X^P(\sigma))$.
If one takes $p=1$ then the Lagrangian (\ref{M2})
describes  the matrix string \cite{Motl,BS,DVV}
in curved background.
For a $0$-brane $X^M=X^M(\tau)$ in the gauge $A_0=0,~g^{0i}=0$
one has
\begin{eqnarray}
\label{M3}
L=-\frac{1}{2}tr[\dot{Y}_i\dot{Y}_jg^{00}g^{ij}
-\frac{1}{2}[Y_i,Y_j][Y_m,Y_n]g^{im}g^{jn}]
\end{eqnarray}
The Lagrangian (\ref{M3}) describes the bosonic part
of M(atrix) theory in curved space. It can be reduced to the bosonic part
of (\ref{I1})
if we take $g^{00}=-1,~g^{ij}=\delta^{ij}$ and $X^M(\tau)=\tau\delta_{M0}$.

Notice that in contrast to the picture with a noncommutative
geometry in the short distance regime in \cite{BFSS}
here in fact we have the classical commutative spacetime
coordinates $X^M$. The matrices $Y_i$ describe the noncommutative
dynamical system in the ordinary classical spacetime.
The corresponding Hamiltonian is
\begin{eqnarray}
\label{M4}
H=\frac{1}{2}tr[P^iP^jg_{ij}
-\frac{1}{2}[Y_i,Y_j][Y_m,Y_n]g^{im}g^{jn}]
\end{eqnarray}
One deals with quantum mechanics in the dependent on time background
$g^{ij}(\tau)=g^{ij}(X(\tau))$.
Now the properties of the matrix
quantum mechanics depend on the choice of the curve $X(\tau)$.
The one-loop effective action for the theory
 (\ref{M3}) with the $p$-brane metric $g_{MN}$ can be
 computed using the background field method by the standard
procedure. One gets corrections to the phase shift $\delta$
obtained in \cite{DKPS,BL,DPS} which are now under consideration.
If one takes a
geodesic near the singularity then generically one gets the creation of
particles (D$0$-branes) \cite{BD}.

If the metric $g_{ij}$ in (\ref{M3}) describes a black hole
then one can apply to M(atrix) theory the known Hawking arguments
\cite{Haw} on the violation of quantum coherence. If one has
M(atrix) theory in flat spacetime (\ref{I1}) then of course
there exists the unitary evolution operator and there is no
a violation of quantum coherence. But in this case simply
there is no  problem for discussion because there are no
black holes in the flat spacetime.

To get the supersymmetric M(atrix) theory or matrix string theory
in curved background
one has to take the dimensional reduction of the supersymmetric
Yang-Mills theory in ten dimensions coupling with supergravity,
\begin{eqnarray}
\label{M7}
L=\frac{1}{4g^2\Phi}tr(F_{MN}^2)-\frac{1}{2}tr(\overline{\chi}
\Gamma^MD_M\chi)-\frac{1}{2\kappa^2}R-\frac{3\kappa^2}{8g^4\Phi^2}
H^2_{MNP}+...
\end{eqnarray}

For the $p$-brane
\begin{eqnarray}
\label{M5}
ds^2=f^{-\frac{1}{2}}(-dt^2+dx_1^2+...+dx_p^2)+f^{\frac{1}{2}}(
dx_{p+1}^2+...+dx_9^2),
\end{eqnarray}
$f=1+\frac{q}{r^{7-p}}$, which is the
BPS-state  we don't expect the Hawking
radiation but there is the spacetime singularity here and there is
back reaction of the gas of D-branes to the metric which
is described by the equations:
\begin{eqnarray}
\label{M6}
R_{\mu\nu}-\frac{1}{2}Rg_{\mu\nu}=<T_{\mu\nu}>
\end{eqnarray}
where $T_{\mu\nu}$ is the energy-momentum tensor of D-branes.

It seems that M(atrix) theory  being quantum mechanics in the curved
spacetime suffers from the well known problems such as non-controllable
spacetime singularities and the violation of quantum coherence.
There is a hope \cite{Vol,DPS} that large $N$ limit gauge theory might
be used to study these problems.

\section{Scattering of D-branes and creation of black holes}

Amati, Ciafaloni and Veneziano \cite{ACV} and 't Hooft \cite{tHo}
have argued that  at extremely high energies
interactions due to gravitational waves will dominate all other
interactions. They conjectured that black holes will occur
in the collision of two light particles at Planckian energies
with small impact parameter. In \cite{ACV,FPVV} the elastic
scattering amplitude  in the eikonal approximation
was found in the form
\begin{eqnarray}
\label{S1}
A(s,t)\propto s\int d^2b e^{iqb} e^{iI_{cl}}
\end{eqnarray}
Here $s$ and $t$ are the Mandelstam variables and $b$ is the impact
parameter. $I_{cl}$ was taken to be the value of the boundary
term for the gravitational action calculated on the sum of two
Aichelburg-Sexl shock  waves,
\begin{eqnarray}
\label{S2}
I_{cl}=Gs\log b^2
\end{eqnarray}
The action $I_{cl}$ is equal to the phase shift $\delta (b,v)$
for the process of the elastic scattering, see \cite{FPVV,AVV}.

One cannot see
the creation of black holes in this approximation. In \cite{AVV} the
following mechanism of the creation of black holes in the process of
collisions of the Planckian energy particles
has been suggested. Each of the
two ultrarelativistic particles generates
a plane gravitational wave. Then
these plane waves collide and produce a
singularity or black hole.  The phase
of the transition amplitude from plane
waves to black holes was calculated as
the value of the action on the corresponding classical solution.

It seems that scattering of D-branes at extremely high energies
and small impact parameter will be similar to the described picture. In
particular two colliding D$0$-branes at
small impact parameter should produce
$p$-branes.

Scattering of D-branes at
large impact parameter has been considered in
\cite{Bac,GKHM,Lif,DKPS,BL}. The $0$-brane metric lifted
to 11 dimensions is
\begin{eqnarray}
\label{S3}
ds^2=dudv+(1+\frac{q}{r^7})du^2+dx_1^2+...+dx^2_9
\end{eqnarray}
It represents a plane-fronted wave moving in the $x_{11}$ direction.
At long distances the gravitational wave can be considered as a plane
wave. Plane wave solutions in supergravity have been considered
in \cite{Gib,Guv,BKO,HT,RT}.
Collision of two plane gravitational waves produce a spacetime that is
locally isometric to an interior of black hole, see \cite{AVV}.
To estimate the amplitude for the creation
of black holes one can use the expression
\begin{eqnarray}
\label{S4}
A\propto \int d^9 b e^{iqb} e^{iI_{cl}}
\end{eqnarray}
where $I_{cl}$ is the value of the boundary term for the gravitational
action of the $D=11$ supergravity \cite{HW} calculated on the solution
describing colliding plane waves \cite{AVV}.
$$~$$
{\bf ACKNOWLEDGMENT}
\vspace{5mm}

The author is grateful to I.Ya. Aref'eva for the useful
discussions.
$$~$$

\end{document}